\documentclass[aps,prl,twocolumn,superscriptaddress,showpacs,floatfix]{revtex4}
\usepackage{graphicx,color}
\usepackage{dcolumn}
\usepackage{hyperref}
\usepackage{amsmath}
\hfuzz=\maxdimen
\usepackage{float}
\begin{document}

\title{Robust Quantum Griffiths Singularity at above 1.5 Kelvin in Nitride Thin Films}


\author{Xiaoni Wang}
\affiliation{State Key Laboratory of Functional Materials for
Informatics, Shanghai Institute of Microsystem and Information
Technology, Chinese Academy of Sciences, Shanghai 200050,
China}\affiliation{University
of Chinese Academy of Science, Beijing 100049, China}

\author{Lijie Wang}
\affiliation{Department of Physics and State Key Laboratory of Surface Physics, Fudan University, Shanghai 200433, China}

\author{Yixin Liu}
\affiliation{State Key Laboratory of Functional Materials for
Informatics, Shanghai Institute of Microsystem and Information
Technology, Chinese Academy of Sciences, Shanghai 200050,
China}\affiliation{University
of Chinese Academy of Science, Beijing 100049, China}

\author{Fan Chen}
\affiliation{State Key Laboratory of Functional Materials for
Informatics, Shanghai Institute of Microsystem and Information
Technology, Chinese Academy of Sciences, Shanghai 200050,
China}\affiliation{University
of Chinese Academy of Science, Beijing 100049, China}

\author{Wanpeng Gao}
\affiliation{State Key Laboratory of Functional Materials for
Informatics, Shanghai Institute of Microsystem and Information
Technology, Chinese Academy of Sciences, Shanghai 200050,
China}\affiliation{University
of Chinese Academy of Science, Beijing 100049, China}

\author{Yu Wu}
\affiliation{State Key Laboratory of Functional Materials for
Informatics, Shanghai Institute of Microsystem and Information
Technology, Chinese Academy of Sciences, Shanghai 200050,
China}

\author{Zulei Xu}
\affiliation{State Key Laboratory of Functional Materials for
Informatics, Shanghai Institute of Microsystem and Information
Technology, Chinese Academy of Sciences, Shanghai 200050,
China}\affiliation{University
of Chinese Academy of Science, Beijing 100049, China}

\author{Wei Peng}
\affiliation{State Key Laboratory of Functional Materials for
Informatics, Shanghai Institute of Microsystem and Information
Technology, Chinese Academy of Sciences, Shanghai 200050,
China}\affiliation{University
of Chinese Academy of Science, Beijing 100049, China}

\author{Zhen Wang}
\affiliation{State Key Laboratory of Functional Materials for
Informatics, Shanghai Institute of Microsystem and Information
Technology, Chinese Academy of Sciences, Shanghai 200050,
China}\affiliation{University
of Chinese Academy of Science, Beijing 100049, China}

\author{Zengfeng Di}
\affiliation{State Key Laboratory of Functional Materials for
Informatics, Shanghai Institute of Microsystem and Information
Technology, Chinese Academy of Sciences, Shanghai 200050,
China}\affiliation{University
of Chinese Academy of Science, Beijing 100049, China}

\author{Wei Li}\email[]{w$_$li@fudan.edu.cn}
\affiliation{Department of Physics and State Key Laboratory of Surface Physics, Fudan University, Shanghai 200433, China}

\author{Gang Mu}
\email[]{mugang@mail.sim.ac.cn} \affiliation{State Key Laboratory of Functional Materials for
Informatics, Shanghai Institute of
Microsystem and Information Technology, Chinese Academy of Sciences,
Shanghai 200050, China}\affiliation{University
of Chinese Academy of Science, Beijing 100049, China}

\author{Zhirong Lin}
\email[]{zrlin@mail.sim.ac.cn} \affiliation{State Key Laboratory of Functional Materials for
Informatics, Shanghai Institute of
Microsystem and Information Technology, Chinese Academy of Sciences,
Shanghai 200050, China}\affiliation{University
of Chinese Academy of Science, Beijing 100049, China}





\begin{abstract}
Quantum Griffiths singularity (QGS), which is closely correlated with the quenched disorder, is characterized by the divergence of the dynamical critical exponent and the presence of activated scaling behavior.
Typically such a quantum phenomenon is rather rare and only observed in extremely low temperatures.
Here we report the experimental observation of a robust QGS in nitride thin films, NbN, which survives in a rather high temperature range. The electrical transport propertied were measured under the magnetic field up to 12 T.
The field induced superconductor-metal transitions
were observed with the continuously changed transition points with the decrease of temperature. The dynamical critical exponent based on the conventional power-law
scaling reveals a divergent trend when approaching the the low temperature limit. Moreover, the temperature and field dependence of sheet resistance can be described by the activated scaling analysis in
the temperature region up to 1.6 K $\leq T \leq$ 4.0 K. We argue that the robustness of QGS in the present system originates from the Pauli paramagnetic effect due to the strong coupling between the magnetic field and
the spin degree of freedom.


\end{abstract}


\maketitle


\section*{1 Introduction}
The phase transition between different ground states, which is accompanied by quantum rather than thermal fluctuations, is called the quantum transition~\cite{Metal-insulator-transitions,Gantmakher_2010}.
For example, the superconductor-insulator
transition (SIT) is typically driven by disorder, magnetic field ($B$), or carrier concentration~\cite{PRL5217,PRB3037,Biscaras2013,Sun2018,PRL257003,TWang2022}.
Such a quantum transition can have a significant influence on the physical performance in finite temperatures
in terms of the quantum fluctuations. In the conventional scenes, only one critical transition point exists and the temperature dependent resistance $R$ near the SIT point can be described by a power-law scaling
determined by the critical exponents ($z$, $\nu$) and the distance from the critical point~\cite{Bollinger2011,Leng2011,Liao2018,Wang2021,Yu2019}.
Recently, it was reported that, in the two-dimensional (2D) or quasi-2D systems with quenched disorder,
a new type of superconductor-metal transition (SMT) emerges. In these systems,
the crossing point in the $R(B)$ curves can reals a systematic variation with temperature and effective critical exponents at each crossing point shows a divergent
behavior when approaching the zero temperature~\cite{Xing2015,PRBShen2016,Xing2017,NCSaito2018,PRBLewellyn2019,Han2020,PRLLiu2021}.

It was interpreted that~\cite{Vojta2010,Spivak2008}, due to the transformation of the vortex lattice
into the vortex glass-like phase in the zero-temperature limit, rare regions of
inhomogeneous superconducting (SC) islands (or puddles) gradually emerge in the high field regime. As a result, the quantum
Griffiths singularity emerges, which manifests an activated scaling behavior in a limited temperature range. It is believed that~\cite{Xing2015} such a vortex glass-like phase is rather delicate against the thermal fluctuation, since
in the high-temperature region, thermal fluctuations will smear the the inhomogeneity caused by quenched disorder. Thus the rare regions of the SC islands could not exist.
In other words, the quantum Griffiths singularity can only be observed in extremely low temperatures, where the impact of quenched disorder overtakes thermal fluctuations.
Although it has been pointed out theoretically that~\cite{Spivak2008} the dominant of spin-related pair-breaking effect will lead to a larger energy associated with the formation of the SC
puddles, it is still unknown concerning the upper limit of temperature that QGS can survive. Especially, there is no related experimental results on this issue.

%

In this study, we conducted an in-depth investigation on the field induced superconductor-metal transition in NbN thin films, whose SC critical temperature $T_c$ has been tuned to below 4 K.
Systematic variation of the crossing points $B_{x}$ and the effective critical exponents $z\nu$ are observed in a finite temperature range. Moreover, the temperature and field dependence of sheet
resistance can be described by the activated dynamical scaling in a relatively wide temperature region up to 4.0 K. Our finding indicates that the features of the quantum Griffiths singularity actually
can survive at a higher temperature than we have expected previously.

\begin{figure}[b]
\includegraphics[width=9cm]{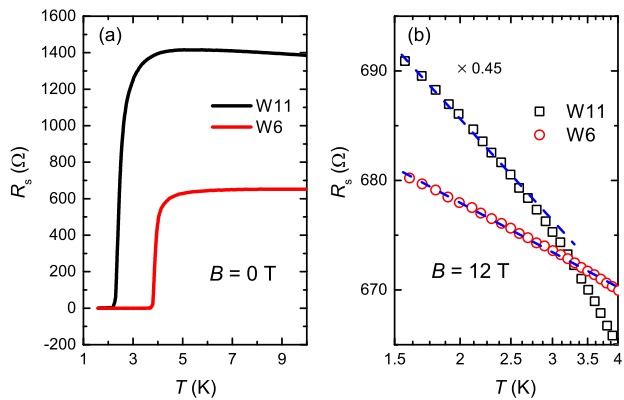}
\centering \caption {Temperature dependence of sheet resistance $R_s$ under the
magnetic fields of 0 T (a) and 12 T (b) for two samples. The semi logarithmic coordinates is used in (b). The blue dashed lines are visual guides.}
\label{fig1}
\end{figure}

\begin{figure*}
\includegraphics[width=13cm]{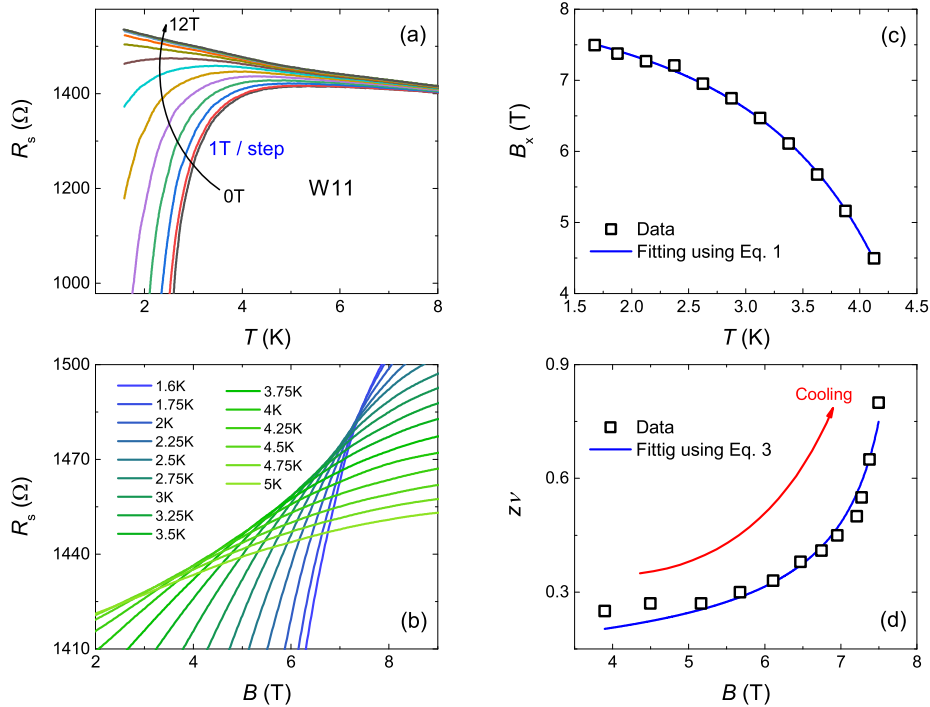}
\caption {(a) Temperature dependence of sheet resistance under magnetic fields up to 12 T for sample W11. (b) Field dependence of sheet resistance at temperatures ranging from 1.6 K to 5 K.
(c) Crossing points from the magnetoresistance isotherms shown in (b). The blue solid line is a fit to the equation (see taxt). (d) Field dependence of the critical exponent $z\nu$
obtained from finite size scaling analysis.  The blue solid line shows the fitting
curve based on the activated scaling law.} \label{fig2}
\end{figure*}

\begin{figure}
\includegraphics[width=8.5cm]{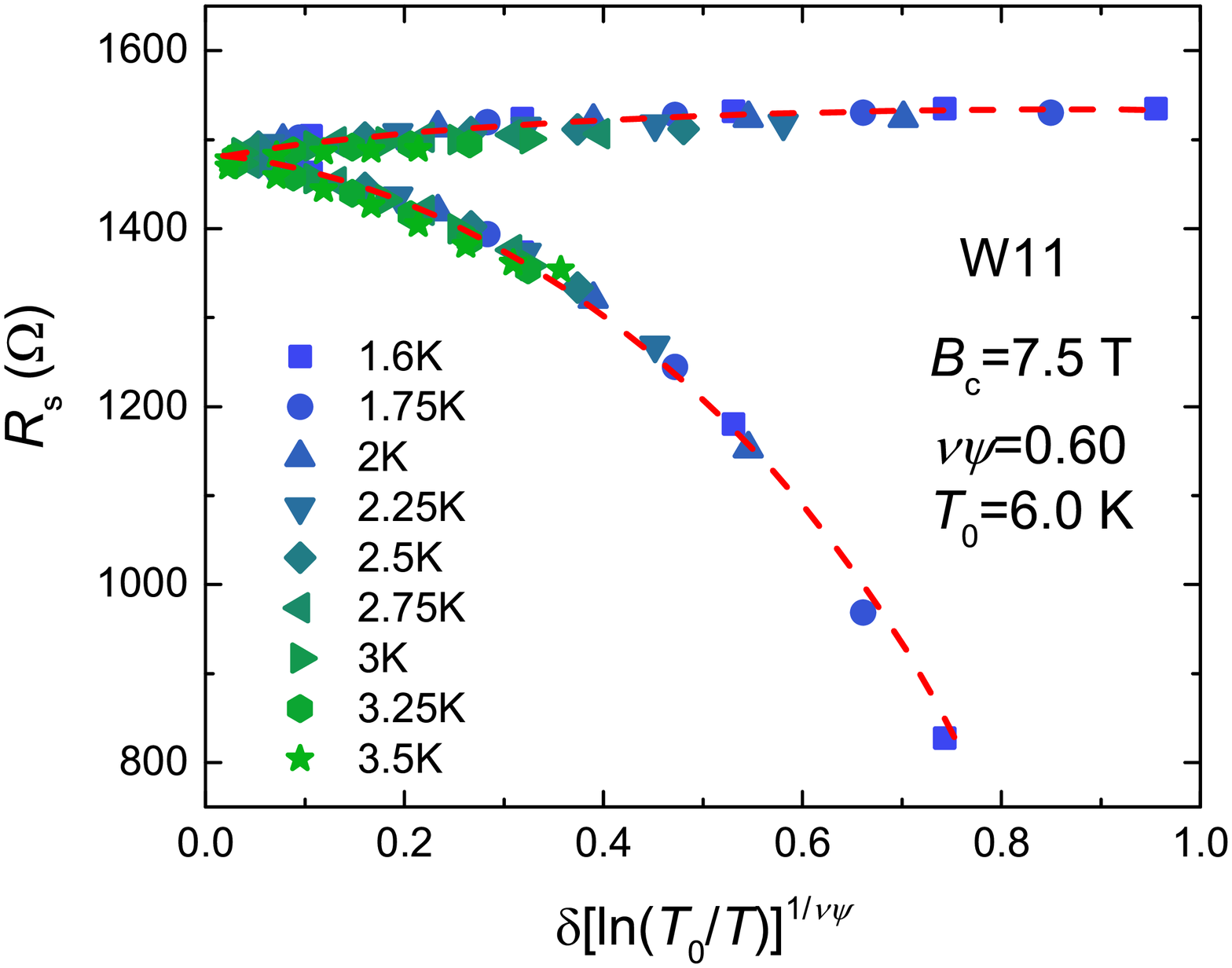}
\caption {Sheet resistance as a function of scaling variable $\delta [ln(T_0/T)]^{1/\nu\psi}$ in the temperature range
1.6 K $\leq T \leq$ 3.5 K. The red dashed lines are visual guides.} \label{fig3}
\end{figure}

\section*{2 Materials and methods}\label{sec:4}
The NbN thin films were deposited using the method of DC reactive magnetron sputtering. The details of the conditions for the film growth have been reported previously~\cite{XNWang2022}.
In the present work, two films (W11 and W6)
grown under different chamber pressures were chosen to conduct the following investigations. The thicknesses of the two films were determined by
the X-ray reflectivity (XRR) measurements to be 25 and 20 nm, respectively. The films were etched into the line shape with the dimension 10$\times$500 $\mu$m$^2$ using reactive ion etching (RIE)
for the electrical transport measurements.

The electrical transport measurements were
performed using a cryostat (Oxford Instruments TeslatronPT cryostat
system) with the field of up to 12 T. The applied
electric current is 0.5 $\mu$A during the transport measurements.
The magnetic field were applied perpendicular to both the film surface and the direction of electrical current.

\section*{3 Results}
The sheet resistance $R_s$ as a function of temperature of the two samples are shown in Fig. 1. Under zero field, as shown in Fig. 1(a), clear SC transitions can be seen in the $R_s$-$T$ curves, with the $T_c$ of
2.5 K and 3.9 K for the samples W11 and W6, respectively. Besides the superconductivity, the magnitude of sheet resistance in the normal state just above $T_c$, $R_N$, also reveals a clear different between the two samples:
$R_N$ = 1400 and 650 $\Omega$ for W11 and W6, respectively. It is notable that the $R_s$-$T$ under a high field of 12 T shows a logarithmic temperature dependence in the low temperature region, see Fig. 1(b). This
feature is the hallmark of the 2D quantum corrected disordered metal, where weak localization paly a crucial role in determining the nature of electrons~\cite{PRBLewellyn2019,Kramer_1993}.

In order to investigate the detailed characteristic of this field induce transition from the SC to metallic state, we measured the temperature dependence of $R_s$ under a series of magnetic fields up to 12 T.
The data of sample W11 are displayed in Fig. 2.
As shown in Fig. 2(a), with the increase of field, the steep descent behavior due the SC transition is suppressed gradually, and eventually turns into a quantum corrected metallic state.
By carrying out a matrix inversion of the temperature-swept data, field dependence of $R_s$ can be generated, which is shown in Fig. 2(b). A series and continuum of crossings in the $R_s$-$B$ curves can be seen, which
spreads out over a range of temperatures (1.6-5 K) and magnetic fields. We extracted the positions of the crossing points $B_{x}$ and showed them in Fig. 2(c) as a function of temperature.
Such an evolution of $B_{x}$ can be simulated using the equation based on the activated scaling~\cite{PRBLewellyn2019}
\begin{equation}
\begin{split}
B_{x}(T)=B_c\times[1-u(ln\frac{T_0}{T})^{-p}], \label{eq:1}
\end{split}
\end{equation}
where $B_c$ is the critical field in the zero-temperature limit, $u$ and $p$ are fitting parameters. The fitting result is presented by the blue solid line in Fig. 2(c).

Similar to the previous reports in other systems~\cite{Xing2015,PRBLewellyn2019,Han2020,PRLLiu2021},
the $R_s$-$T$ data in Fig. 2(b) in each small temperature intervals can be analyzed using the conventional power-law scaling
\begin{equation}
\begin{split}
R_s(\delta_x,T)=F(\delta_x T^{-1/z\nu}), \label{eq:2}
\end{split}
\end{equation}
where $\delta_x=|B-B_x|/B_x$ is the normalized distance from the effective critical field $B_x$ in each small temperature ranges, $\nu$ is the correlation-length
exponent, $z$ is the dynamical-scaling exponent, and $F$ is the scaling function with $F(0)$ = 1. The results of the scaling using Eq. 2
are shown in the Supplementary Material (SM). The obtained critical exponents $z\nu$ as a function of field are shown in Fig. 2(d). With the decrease of temperature, the value of $z\nu$
reveals an upward trend with an increasing steepness. Such a divergent tendency is simulated using the equation based on the activated scaling
\begin{equation}
\begin{split}
z\nu = C\times|B-B_c|^{-0.6}, \label{eq:3}
\end{split}
\end{equation}
where $C$ is the fitting parameter. The power exponent is set as the value ($\sim$0.6) used by other groups~\cite{Xing2015,PRLLiu2021}.
As shown by the blue solid line, the experimental data, especially the divergent tendency in the low
temperature region, can be well described by Eq. 3.

\begin{figure*}
\includegraphics[width=13cm]{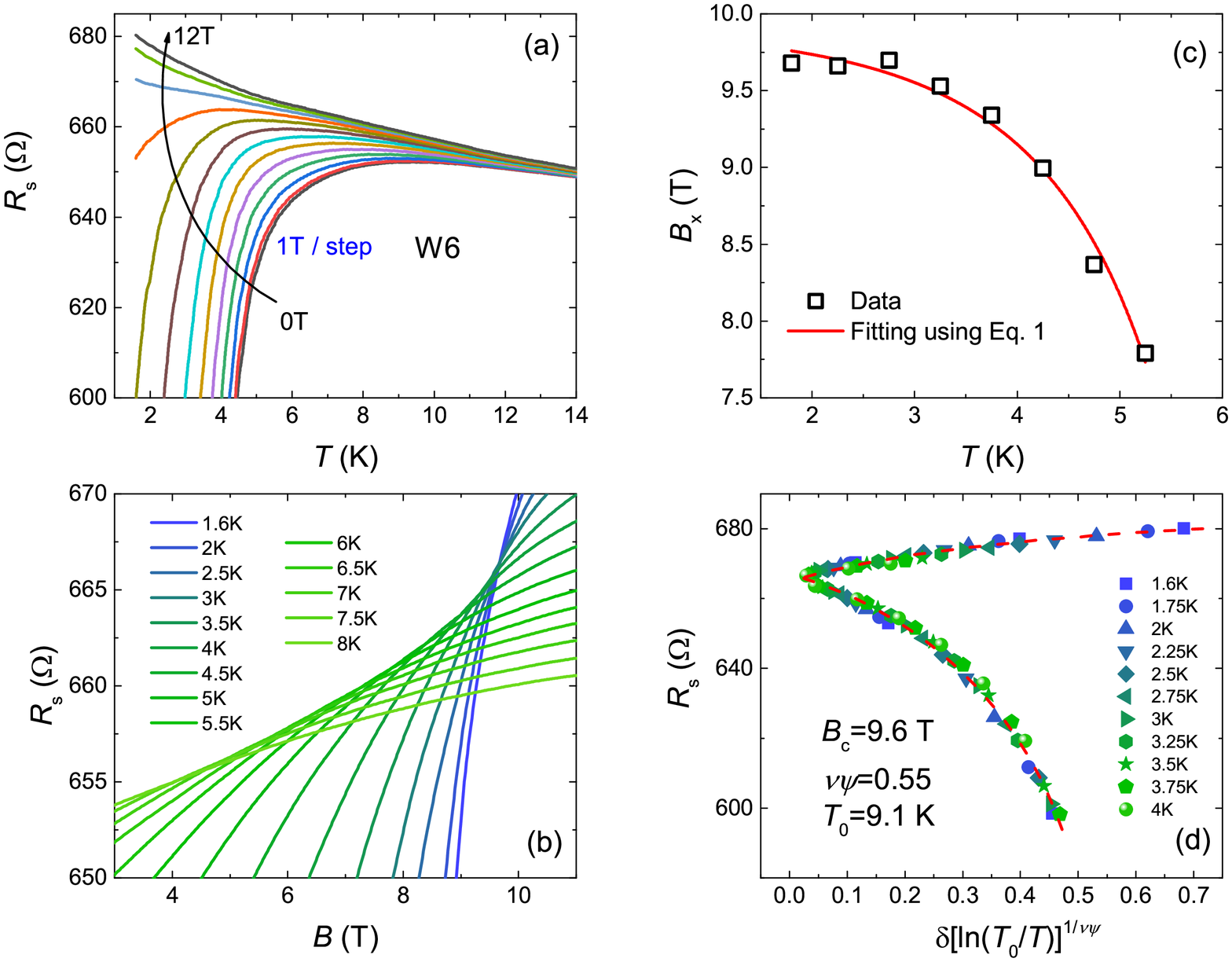}
\caption {(a) Temperature dependence of sheet resistance under magnetic fields up to 12 T for sample W6. (b) Field dependence of sheet resistance at temperatures ranging from 1.6 K to 8 K.
(c) Crossing points from the magnetoresistance isotherms shown in (b). The red solid line is a fit to the equation (see taxt).
(d) Sheet resistance as a function of scaling variable $\delta [ln(T_0/T)]^{1/\nu\psi}$ in the temperature range
1.6 K $\leq T \leq$ 4 K. The red dashed lines are visual guides.} \label{fig4}
\end{figure*}

Theoretically, quantum Griffiths singularity is closely related with an infinite-randomness critical point. Strictly speaking, the quantum SMT governed by such an infinite-randomness fixed point could not be described
by the conventional power-law scaling. Recently, the activated dynamical scaling was proposed~\cite{PRBLewellyn2019} in the following form
\begin{equation}
\begin{split}
R_s(\delta,ln\frac{T_0}{T})=\Phi[\delta (ln\frac{T_0}{T})^{1/\nu\psi}]. \label{eq:4}
\end{split}
\end{equation}
Here $\delta=|B-B_c|/B_c$ is the normalized distance from the critical field $B_c$, exponent $\psi$ is the tunneling exponent, and $T_0$ is a microscopic temperature scale. $\Phi$ is the function of the activated dynamical scaling.
According to this equation,
there is only one single crossing point in magnetic field, which does not account for the temperature dependence of the crossing fields in the experimental data.
In Fig. 3, we show the results of the activated dynamical scaling according to Eq. 4. All the experimental data in the temperature range 1.6 K $\leq T \leq$ 3.5 K collapse into a single
curve, indicating the rationality of the activated dynamical scaling in describing our data. From this scaling, the critical
exponent $\nu\psi$ = 0.60 and critical field $B_c$ = 7.50 T were obtained.
The value of $\nu\psi$ obtained here is slightly smaller than that reported in InO$_x$ films ($\sim$ 0.62)~\cite{PRBLewellyn2019}.

Sample W6, with a relatively higher $T_c$ and lower $R_N$, also reveals the field induced SMT, as shown in Fig. 4(a). Similar to that observed in sample W11, the crossing points in the $R_s$-$B$ curves
experience a series of evolution with temperature, which tends to be saturated at above 9 T as the temperature goes towards lower temperatures (see Fig. 4(b)). The values of the crossing fields $B_x$ are recorded
in Fig. 4(c), from which the saturated tendency can be seen more clearly. Again the $B_x$-$T$ data are simulated using Eq. 1 as revealed by the red solid line in Fig. 4(c). The temperature and field dependence of
sheet resistance in the temperature range 1.6 K $\leq T \leq$ 4.0 K are analyzed using the activated dynamical scaling, which are shown in Fig. 4(d). The scaling parameters are $B_c$ = 9.6 T, $\nu\psi$ = 0.55, and $T_0$ = 9.1 K. These parameters of the two samples are summarized
in Table 1 to have a comparison.

\begin{table}
\centering
\caption{Important parameters of the two samples.}
\begin{tabular}
{p{1.2cm}<{\centering}p{1.2cm}<{\centering}p{1.2cm}<{\centering}p{1.2cm}<{\centering}p{1.2cm}<{\centering}p{1.2cm}<{\centering}}\hline \hline
   Name    &   $T_c$   &  $R_N$ & $B_c$ & $\nu\psi$  & $T_0$   \\
             &   (K)   & ($\Omega$) & (T)  &  & (K) \\
\hline
W11    & 2.5      & 1400  &  7.50  & 0.60 & 6.0  \\
W6      &3.9      & 650   &  9.60  & 0.55 & 9.1    \\
\hline
\hline
\end{tabular}
\label{tab.1}
\end{table}

\section*{4 Discussion}
The adjustability in a large range of the physical properties of NbN films~\cite{Oya1974,Wang1996,XNWang2022} provides convenience to broaden the extension of ground state in investigating the quantum phenomena,
like quantum Griffiths singularity in field induced SMT. In this work, the sheet resistance in the normal state is tuned from 1400 $\Omega$ to 650 $\Omega$. Our results suggest that, the quantum Griffiths singularity
can be observed in the system with $R_N$ being suppressed to half. In the case of PdTe$_2$ films~\cite{PRLLiu2021}, the features of quantum Griffiths singularity has disappeared under perpendicular field when $R_N$ is reduced
to below 200 $\Omega$ by increasing the film thickness. Finding and determining the critical point of where QGS disappears is of great significance for revealing the essence of this quantum phenomenon. Based on the NbN film,
we can systematically adjust its normal state resistance without changing the film thickness, so that the boundary of the QGS state can be accurately determined without changing other factors. This will be the next step of our work.

Another important issue worth our attention is the temperature range where QGS occurs. Based on general understanding~\cite{Xing2015}, QGS is rather delicate against the thermal fluctuation,
since the the inhomogeneity caused by quenched disorder
can be smeared by the thermal fluctuations in the high-temperature region. In this case, the rare regions of the SC islands (puddles) could hardly survive. Indeed, previously the signatures of QGS were typically
detected in the extremely low temperatures. In our work, however, it is found that the activated dynamical scaling can described the electrical transport data in the temperature range 1.6 K $\leq T \leq$ 3.5 K  (see Fig. 3) or
1.6 K $\leq T \leq$ 4.0 K (see Fig. 4(d)).

Theoretically, the size, distance, and concentration of the the SC puddles are important parameters in controlling the performance of the QGS~\cite{Spivak2008}.
These factors determine the coupling strength between optimal puddles. Moreover, in the magnetic field induced SMT, the modes of coupling between the applied field and the SC state can also play an important role, which
may give a different microscopic physics responsible for the emergence of the puddle state.
It was pointed out that~\cite{Spivak2008}, compared with the case that the field couples primarily to the electron's orbital motion, the coupling to the electron spin (Zeeman coupling) in the absence of spin-orbit coupling
can lead to a larger energy associated with the formation of puddles. In this case, the puddle state should be more robust. In our previous work~\cite{XNWang2022},
we have uncovered a clear Pauli paramagnetic effect in the present NbN system,
which emphasizes the significant influence of the coupling between the magnetic field and spin degree of freedom. It seems that this is the most probable reason why we have observed such a robust QGS state in NbN films.
We expect to observe similar behaviors in more systems with a strong Pauli paramagnetic effect in the future.

\section*{5 Conclusions}
In summary, we studied the field induced superconductor-metal transition in the NbN thin films. The quantum Griffiths singularity, in terms of the systematical evolution of the crossing point $B_x$ in the $R_s$-$B$ curves
and the applicability of the activated scaling analysis to the transport data, was observed in two samples with different $T_c$ and $R_N$. Significantly, activated scaling can be valid in the temperature rang as high as
1.6 K -- 4.0 K, revealing the robustness of the QGS in the present system. Our analysis indicates that the strong Pauli paramagnetic effect in NbN films plays a key role in enhancing the stability of the puddle state, which
is the precondition for the formation of QGS.


\emph{
This work is
supported by the National
Natural Science Foundation of China (No. 92065116), the Shanghai Technology Innovation Action Plan Integrated Circuit Technology Support Program (No. 22DZ1100200), and
the Key-Area Research and Development Program of Guangdong Province, China (No. 2020B0303030002). The authors would like to thank all the staff at the
Superconducting Electronics Facility (SELF) for their
assistance.}





\begin{thebibliography}{26}
\expandafter\ifx\csname natexlab\endcsname\relax\def\natexlab#1{#1}\fi
\expandafter\ifx\csname bibnamefont\endcsname\relax
  \def\bibnamefont#1{#1}\fi
\expandafter\ifx\csname bibfnamefont\endcsname\relax
  \def\bibfnamefont#1{#1}\fi
\expandafter\ifx\csname citenamefont\endcsname\relax
  \def\citenamefont#1{#1}\fi
\expandafter\ifx\csname url\endcsname\relax
  \def\url#1{\texttt{#1}}\fi
\expandafter\ifx\csname urlprefix\endcsname\relax\def\urlprefix{URL }\fi
\providecommand{\bibinfo}[2]{#2}
\providecommand{\eprint}[2][]{\url{#2}}

\bibitem[{\citenamefont{Imada et~al.}(1998)\citenamefont{Imada, Fujimori, and
  Tokura}}]{Metal-insulator-transitions}
\bibinfo{author}{\bibfnamefont{M.}~\bibnamefont{Imada}},
  \bibinfo{author}{\bibfnamefont{A.}~\bibnamefont{Fujimori}}, \bibnamefont{and}
  \bibinfo{author}{\bibfnamefont{Y.}~\bibnamefont{Tokura}},
  \bibinfo{journal}{Rev. Mod. Phys.} \textbf{\bibinfo{volume}{70}},
  \bibinfo{pages}{1039} (\bibinfo{year}{1998}).

\bibitem[{\citenamefont{Gantmakher and Dolgopolov}(2010)}]{Gantmakher_2010}
\bibinfo{author}{\bibfnamefont{V.~F.} \bibnamefont{Gantmakher}}
  \bibnamefont{and} \bibinfo{author}{\bibfnamefont{V.~T.}
  \bibnamefont{Dolgopolov}}, \bibinfo{journal}{Physics-Uspekhi}
  \textbf{\bibinfo{volume}{53}}, \bibinfo{pages}{1} (\bibinfo{year}{2010}).

\bibitem[{\citenamefont{Markovi\ifmmode~\acute{c}\else \'{c}\fi{}
  et~al.}(1998)\citenamefont{Markovi\ifmmode~\acute{c}\else \'{c}\fi{},
  Christiansen, and Goldman}}]{PRL5217}
\bibinfo{author}{\bibfnamefont{N.}~\bibnamefont{Markovi\ifmmode~\acute{c}\else
  \'{c}\fi{}}}, \bibinfo{author}{\bibfnamefont{C.}~\bibnamefont{Christiansen}},
  \bibnamefont{and} \bibinfo{author}{\bibfnamefont{A.~M.}
  \bibnamefont{Goldman}}, \bibinfo{journal}{Phys. Rev. Lett.}
  \textbf{\bibinfo{volume}{81}}, \bibinfo{pages}{5217} (\bibinfo{year}{1998}).

\bibitem[{\citenamefont{Yazdani and Kapitulnik}(1995)}]{PRB3037}
\bibinfo{author}{\bibfnamefont{A.}~\bibnamefont{Yazdani}} \bibnamefont{and}
  \bibinfo{author}{\bibfnamefont{A.}~\bibnamefont{Kapitulnik}},
  \bibinfo{journal}{Phys. Rev. Lett.} \textbf{\bibinfo{volume}{74}},
  \bibinfo{pages}{3037} (\bibinfo{year}{1995}).

\bibitem[{\citenamefont{Biscaras et~al.}(2013)\citenamefont{Biscaras, Bergeal,
  Hurand, Feuillet-Palma, Rastogi, Budhani, Grilli, Caprara, and
  Lesueur}}]{Biscaras2013}
\bibinfo{author}{\bibfnamefont{J.}~\bibnamefont{Biscaras}},
  \bibinfo{author}{\bibfnamefont{N.}~\bibnamefont{Bergeal}},
  \bibinfo{author}{\bibfnamefont{S.}~\bibnamefont{Hurand}},
  \bibinfo{author}{\bibfnamefont{C.}~\bibnamefont{Feuillet-Palma}},
  \bibinfo{author}{\bibfnamefont{A.}~\bibnamefont{Rastogi}},
  \bibinfo{author}{\bibfnamefont{R.~C.} \bibnamefont{Budhani}},
  \bibinfo{author}{\bibfnamefont{M.}~\bibnamefont{Grilli}},
  \bibinfo{author}{\bibfnamefont{S.}~\bibnamefont{Caprara}}, \bibnamefont{and}
  \bibinfo{author}{\bibfnamefont{J.}~\bibnamefont{Lesueur}},
  \bibinfo{journal}{Nat. Mater.} \textbf{\bibinfo{volume}{12}},
  \bibinfo{pages}{542} (\bibinfo{year}{2013}).

\bibitem[{\citenamefont{Sun et~al.}(2018)\citenamefont{Sun, Xiao, Zhang, Xue,
  Mei, Xie, Hu, Di, and Wang}}]{Sun2018}
\bibinfo{author}{\bibfnamefont{Y.}~\bibnamefont{Sun}},
  \bibinfo{author}{\bibfnamefont{H.}~\bibnamefont{Xiao}},
  \bibinfo{author}{\bibfnamefont{M.}~\bibnamefont{Zhang}},
  \bibinfo{author}{\bibfnamefont{Z.}~\bibnamefont{Xue}},
  \bibinfo{author}{\bibfnamefont{Y.}~\bibnamefont{Mei}},
  \bibinfo{author}{\bibfnamefont{X.}~\bibnamefont{Xie}},
  \bibinfo{author}{\bibfnamefont{T.}~\bibnamefont{Hu}},
  \bibinfo{author}{\bibfnamefont{Z.}~\bibnamefont{Di}}, \bibnamefont{and}
  \bibinfo{author}{\bibfnamefont{X.}~\bibnamefont{Wang}},
  \bibinfo{journal}{Nat. Commun.} \textbf{\bibinfo{volume}{9}},
  \bibinfo{pages}{2159} (\bibinfo{year}{2018}).

\bibitem[{\citenamefont{Schneider et~al.}(2012)\citenamefont{Schneider,
  Zaitsev, Fuchs, and v.~L\"ohneysen}}]{PRL257003}
\bibinfo{author}{\bibfnamefont{R.}~\bibnamefont{Schneider}},
  \bibinfo{author}{\bibfnamefont{A.~G.} \bibnamefont{Zaitsev}},
  \bibinfo{author}{\bibfnamefont{D.}~\bibnamefont{Fuchs}}, \bibnamefont{and}
  \bibinfo{author}{\bibfnamefont{H.}~\bibnamefont{v.~L\"ohneysen}},
  \bibinfo{journal}{Phys. Rev. Lett.} \textbf{\bibinfo{volume}{108}},
  \bibinfo{pages}{257003} (\bibinfo{year}{2012}).

\bibitem[{\citenamefont{Wang et~al.}(2022{\natexlab{a}})\citenamefont{Wang, Yu,
  Liu, Gu, Peng, Di, Jiang, and Mu}}]{TWang2022}
\bibinfo{author}{\bibfnamefont{T.}~\bibnamefont{Wang}},
  \bibinfo{author}{\bibfnamefont{A.}~\bibnamefont{Yu}},
  \bibinfo{author}{\bibfnamefont{Y.}~\bibnamefont{Liu}},
  \bibinfo{author}{\bibfnamefont{G.}~\bibnamefont{Gu}},
  \bibinfo{author}{\bibfnamefont{W.}~\bibnamefont{Peng}},
  \bibinfo{author}{\bibfnamefont{Z.}~\bibnamefont{Di}},
  \bibinfo{author}{\bibfnamefont{D.}~\bibnamefont{Jiang}}, \bibnamefont{and}
  \bibinfo{author}{\bibfnamefont{G.}~\bibnamefont{Mu}}, \bibinfo{journal}{Phys.
  Rev. B} \textbf{\bibinfo{volume}{106}}, \bibinfo{pages}{104509}
  (\bibinfo{year}{2022}{\natexlab{a}}).

\bibitem[{\citenamefont{Bollinger et~al.}(2011)\citenamefont{Bollinger, Dubuis,
  Yoon, Pavuna, Misewich, and Bo\u{z}ovi\'{c}}}]{Bollinger2011}
\bibinfo{author}{\bibfnamefont{A.~T.} \bibnamefont{Bollinger}},
  \bibinfo{author}{\bibfnamefont{G.}~\bibnamefont{Dubuis}},
  \bibinfo{author}{\bibfnamefont{J.}~\bibnamefont{Yoon}},
  \bibinfo{author}{\bibfnamefont{D.}~\bibnamefont{Pavuna}},
  \bibinfo{author}{\bibfnamefont{J.}~\bibnamefont{Misewich}}, \bibnamefont{and}
  \bibinfo{author}{\bibfnamefont{I.}~\bibnamefont{Bo\u{z}ovi\'{c}}},
  \bibinfo{journal}{Nature} \textbf{\bibinfo{volume}{365}},
  \bibinfo{pages}{458} (\bibinfo{year}{2011}).

\bibitem[{\citenamefont{Leng et~al.}(2011)\citenamefont{Leng,
  Garcia-Barriocanal, Bose, Lee, and Goldman}}]{Leng2011}
\bibinfo{author}{\bibfnamefont{X.}~\bibnamefont{Leng}},
  \bibinfo{author}{\bibfnamefont{J.}~\bibnamefont{Garcia-Barriocanal}},
  \bibinfo{author}{\bibfnamefont{S.}~\bibnamefont{Bose}},
  \bibinfo{author}{\bibfnamefont{Y.}~\bibnamefont{Lee}}, \bibnamefont{and}
  \bibinfo{author}{\bibfnamefont{A.~M.} \bibnamefont{Goldman}},
  \bibinfo{journal}{Phys. Rev. Lett.} \textbf{\bibinfo{volume}{107}},
  \bibinfo{pages}{027001} (\bibinfo{year}{2011}).

\bibitem[{\citenamefont{Liao et~al.}(2018)\citenamefont{Liao, Zhu, Zhang,
  Zhong, Schneeloch, Gu, Jiang, Zhang, Ma, and Xue}}]{Liao2018}
\bibinfo{author}{\bibfnamefont{M.}~\bibnamefont{Liao}},
  \bibinfo{author}{\bibfnamefont{Y.}~\bibnamefont{Zhu}},
  \bibinfo{author}{\bibfnamefont{J.}~\bibnamefont{Zhang}},
  \bibinfo{author}{\bibfnamefont{R.}~\bibnamefont{Zhong}},
  \bibinfo{author}{\bibfnamefont{J.}~\bibnamefont{Schneeloch}},
  \bibinfo{author}{\bibfnamefont{G.}~\bibnamefont{Gu}},
  \bibinfo{author}{\bibfnamefont{K.}~\bibnamefont{Jiang}},
  \bibinfo{author}{\bibfnamefont{D.}~\bibnamefont{Zhang}},
  \bibinfo{author}{\bibfnamefont{X.}~\bibnamefont{Ma}}, \bibnamefont{and}
  \bibinfo{author}{\bibfnamefont{Q.~K.} \bibnamefont{Xue}},
  \bibinfo{journal}{Nano Lett.} \textbf{\bibinfo{volume}{18}},
  \bibinfo{pages}{5660} (\bibinfo{year}{2018}).

\bibitem[{\citenamefont{Wang et~al.}(2021)\citenamefont{Wang, Biscaras, Erb,
  and Shukla}}]{Wang2021}
\bibinfo{author}{\bibfnamefont{F.}~\bibnamefont{Wang}},
  \bibinfo{author}{\bibfnamefont{J.}~\bibnamefont{Biscaras}},
  \bibinfo{author}{\bibfnamefont{A.}~\bibnamefont{Erb}}, \bibnamefont{and}
  \bibinfo{author}{\bibfnamefont{A.}~\bibnamefont{Shukla}},
  \bibinfo{journal}{Nat. Commun.} \textbf{\bibinfo{volume}{12}},
  \bibinfo{pages}{2926} (\bibinfo{year}{2021}).

\bibitem[{\citenamefont{Yu et~al.}(2019)\citenamefont{Yu, Ma, Cai, Zhong, Ye,
  Shen, Gu, Chen, and Zhang}}]{Yu2019}
\bibinfo{author}{\bibfnamefont{Y.}~\bibnamefont{Yu}},
  \bibinfo{author}{\bibfnamefont{L.}~\bibnamefont{Ma}},
  \bibinfo{author}{\bibfnamefont{P.}~\bibnamefont{Cai}},
  \bibinfo{author}{\bibfnamefont{R.}~\bibnamefont{Zhong}},
  \bibinfo{author}{\bibfnamefont{C.}~\bibnamefont{Ye}},
  \bibinfo{author}{\bibfnamefont{J.}~\bibnamefont{Shen}},
  \bibinfo{author}{\bibfnamefont{G.}~\bibnamefont{Gu}},
  \bibinfo{author}{\bibfnamefont{X.~H.} \bibnamefont{Chen}}, \bibnamefont{and}
  \bibinfo{author}{\bibfnamefont{Y.}~\bibnamefont{Zhang}},
  \bibinfo{journal}{Nature} \textbf{\bibinfo{volume}{575}},
  \bibinfo{pages}{156} (\bibinfo{year}{2019}).

\bibitem[{\citenamefont{Xing et~al.}(2015)\citenamefont{Xing, Zhang, Fu, Liu,
  Sun, Peng, Wang, Lin, Ma, Xue et~al.}}]{Xing2015}
\bibinfo{author}{\bibfnamefont{Y.}~\bibnamefont{Xing}},
  \bibinfo{author}{\bibfnamefont{H.-M.} \bibnamefont{Zhang}},
  \bibinfo{author}{\bibfnamefont{H.-L.} \bibnamefont{Fu}},
  \bibinfo{author}{\bibfnamefont{H.}~\bibnamefont{Liu}},
  \bibinfo{author}{\bibfnamefont{Y.}~\bibnamefont{Sun}},
  \bibinfo{author}{\bibfnamefont{J.-P.} \bibnamefont{Peng}},
  \bibinfo{author}{\bibfnamefont{F.}~\bibnamefont{Wang}},
  \bibinfo{author}{\bibfnamefont{X.}~\bibnamefont{Lin}},
  \bibinfo{author}{\bibfnamefont{X.-C.} \bibnamefont{Ma}},
  \bibinfo{author}{\bibfnamefont{Q.-K.} \bibnamefont{Xue}},
  \bibnamefont{et~al.}, \bibinfo{journal}{Science}
  \textbf{\bibinfo{volume}{350}}, \bibinfo{pages}{542} (\bibinfo{year}{2015}).

\bibitem[{\citenamefont{Shen et~al.}(2016)\citenamefont{Shen, Xing, Wang, Liu,
  Fu, Zhang, He, Xie, Lin, Nie et~al.}}]{PRBShen2016}
\bibinfo{author}{\bibfnamefont{S.}~\bibnamefont{Shen}},
  \bibinfo{author}{\bibfnamefont{Y.}~\bibnamefont{Xing}},
  \bibinfo{author}{\bibfnamefont{P.}~\bibnamefont{Wang}},
  \bibinfo{author}{\bibfnamefont{H.}~\bibnamefont{Liu}},
  \bibinfo{author}{\bibfnamefont{H.}~\bibnamefont{Fu}},
  \bibinfo{author}{\bibfnamefont{Y.}~\bibnamefont{Zhang}},
  \bibinfo{author}{\bibfnamefont{L.}~\bibnamefont{He}},
  \bibinfo{author}{\bibfnamefont{X.~C.} \bibnamefont{Xie}},
  \bibinfo{author}{\bibfnamefont{X.}~\bibnamefont{Lin}},
  \bibinfo{author}{\bibfnamefont{J.}~\bibnamefont{Nie}}, \bibnamefont{et~al.},
  \bibinfo{journal}{Phys. Rev. B} \textbf{\bibinfo{volume}{94}},
  \bibinfo{pages}{144517} (\bibinfo{year}{2016}).

\bibitem[{\citenamefont{Xing et~al.}(2017)\citenamefont{Xing, Zhao, Shan,
  Zheng, Zhang, Fu, Liu, Tian, Xi, Liu et~al.}}]{Xing2017}
\bibinfo{author}{\bibfnamefont{Y.}~\bibnamefont{Xing}},
  \bibinfo{author}{\bibfnamefont{K.}~\bibnamefont{Zhao}},
  \bibinfo{author}{\bibfnamefont{P.}~\bibnamefont{Shan}},
  \bibinfo{author}{\bibfnamefont{F.}~\bibnamefont{Zheng}},
  \bibinfo{author}{\bibfnamefont{Y.}~\bibnamefont{Zhang}},
  \bibinfo{author}{\bibfnamefont{H.}~\bibnamefont{Fu}},
  \bibinfo{author}{\bibfnamefont{Y.}~\bibnamefont{Liu}},
  \bibinfo{author}{\bibfnamefont{M.}~\bibnamefont{Tian}},
  \bibinfo{author}{\bibfnamefont{C.}~\bibnamefont{Xi}},
  \bibinfo{author}{\bibfnamefont{H.}~\bibnamefont{Liu}}, \bibnamefont{et~al.},
  \bibinfo{journal}{Nano Letters} \textbf{\bibinfo{volume}{17}},
  \bibinfo{pages}{6802} (\bibinfo{year}{2017}).

\bibitem[{\citenamefont{Saito et~al.}(2018)\citenamefont{Saito, Nojima, and
  Iwasa}}]{NCSaito2018}
\bibinfo{author}{\bibfnamefont{Y.}~\bibnamefont{Saito}},
  \bibinfo{author}{\bibfnamefont{T.}~\bibnamefont{Nojima}}, \bibnamefont{and}
  \bibinfo{author}{\bibfnamefont{Y.}~\bibnamefont{Iwasa}},
  \bibinfo{journal}{Nat. Commun.} \textbf{\bibinfo{volume}{9}},
  \bibinfo{pages}{778} (\bibinfo{year}{2018}).

\bibitem[{\citenamefont{Lewellyn et~al.}(2019)\citenamefont{Lewellyn, Percher,
  Nelson, Garcia-Barriocanal, Volotsenko, Frydman, Vojta, and
  Goldman}}]{PRBLewellyn2019}
\bibinfo{author}{\bibfnamefont{N.~A.} \bibnamefont{Lewellyn}},
  \bibinfo{author}{\bibfnamefont{I.~M.} \bibnamefont{Percher}},
  \bibinfo{author}{\bibfnamefont{J.}~\bibnamefont{Nelson}},
  \bibinfo{author}{\bibfnamefont{J.}~\bibnamefont{Garcia-Barriocanal}},
  \bibinfo{author}{\bibfnamefont{I.}~\bibnamefont{Volotsenko}},
  \bibinfo{author}{\bibfnamefont{A.}~\bibnamefont{Frydman}},
  \bibinfo{author}{\bibfnamefont{T.}~\bibnamefont{Vojta}}, \bibnamefont{and}
  \bibinfo{author}{\bibfnamefont{A.~M.} \bibnamefont{Goldman}},
  \bibinfo{journal}{Phys. Rev. B} \textbf{\bibinfo{volume}{99}},
  \bibinfo{pages}{054515} (\bibinfo{year}{2019}).

\bibitem[{\citenamefont{Han et~al.}(2020)\citenamefont{Han, Wu, Xiao, Zhang,
  Gao, Liu, Wang, Hu, Xie, and Di}}]{Han2020}
\bibinfo{author}{\bibfnamefont{X.}~\bibnamefont{Han}},
  \bibinfo{author}{\bibfnamefont{Y.}~\bibnamefont{Wu}},
  \bibinfo{author}{\bibfnamefont{H.}~\bibnamefont{Xiao}},
  \bibinfo{author}{\bibfnamefont{M.}~\bibnamefont{Zhang}},
  \bibinfo{author}{\bibfnamefont{M.}~\bibnamefont{Gao}},
  \bibinfo{author}{\bibfnamefont{Y.}~\bibnamefont{Liu}},
  \bibinfo{author}{\bibfnamefont{J.}~\bibnamefont{Wang}},
  \bibinfo{author}{\bibfnamefont{T.}~\bibnamefont{Hu}},
  \bibinfo{author}{\bibfnamefont{X.}~\bibnamefont{Xie}}, \bibnamefont{and}
  \bibinfo{author}{\bibfnamefont{Z.}~\bibnamefont{Di}}, \bibinfo{journal}{Adv.
  Sci.} \textbf{\bibinfo{volume}{7}}, \bibinfo{pages}{1902849}
  (\bibinfo{year}{2020}).

\bibitem[{\citenamefont{Liu et~al.}(2021)\citenamefont{Liu, Qi, Fang, Sun, Liu,
  Liu, Qi, Xing, Liu, Lin et~al.}}]{PRLLiu2021}
\bibinfo{author}{\bibfnamefont{Y.}~\bibnamefont{Liu}},
  \bibinfo{author}{\bibfnamefont{S.}~\bibnamefont{Qi}},
  \bibinfo{author}{\bibfnamefont{J.}~\bibnamefont{Fang}},
  \bibinfo{author}{\bibfnamefont{J.}~\bibnamefont{Sun}},
  \bibinfo{author}{\bibfnamefont{C.}~\bibnamefont{Liu}},
  \bibinfo{author}{\bibfnamefont{Y.}~\bibnamefont{Liu}},
  \bibinfo{author}{\bibfnamefont{J.}~\bibnamefont{Qi}},
  \bibinfo{author}{\bibfnamefont{Y.}~\bibnamefont{Xing}},
  \bibinfo{author}{\bibfnamefont{H.}~\bibnamefont{Liu}},
  \bibinfo{author}{\bibfnamefont{X.}~\bibnamefont{Lin}}, \bibnamefont{et~al.},
  \bibinfo{journal}{Phys. Rev. Lett.} \textbf{\bibinfo{volume}{127}},
  \bibinfo{pages}{137001} (\bibinfo{year}{2021}).

\bibitem[{\citenamefont{Vojta}(2010)}]{Vojta2010}
\bibinfo{author}{\bibfnamefont{T.}~\bibnamefont{Vojta}}, \bibinfo{journal}{J.
  Low Temp. Phys.} \textbf{\bibinfo{volume}{161}}, \bibinfo{pages}{299}
  (\bibinfo{year}{2010}).

\bibitem[{\citenamefont{Spivak et~al.}(2008)\citenamefont{Spivak, Oreto, and
  Kivelson}}]{Spivak2008}
\bibinfo{author}{\bibfnamefont{B.}~\bibnamefont{Spivak}},
  \bibinfo{author}{\bibfnamefont{P.}~\bibnamefont{Oreto}}, \bibnamefont{and}
  \bibinfo{author}{\bibfnamefont{S.~A.} \bibnamefont{Kivelson}},
  \bibinfo{journal}{Phys. Rev. B} \textbf{\bibinfo{volume}{77}},
  \bibinfo{pages}{214523} (\bibinfo{year}{2008}).

\bibitem[{\citenamefont{Wang et~al.}(2022{\natexlab{b}})\citenamefont{Wang,
  Wang, Liu, Gao, Wu, Xu, Jin, Zhang, Peng, Wang et~al.}}]{XNWang2022}
\bibinfo{author}{\bibfnamefont{X.}~\bibnamefont{Wang}},
  \bibinfo{author}{\bibfnamefont{L.}~\bibnamefont{Wang}},
  \bibinfo{author}{\bibfnamefont{Y.}~\bibnamefont{Liu}},
  \bibinfo{author}{\bibfnamefont{W.}~\bibnamefont{Gao}},
  \bibinfo{author}{\bibfnamefont{Y.}~\bibnamefont{Wu}},
  \bibinfo{author}{\bibfnamefont{Z.}~\bibnamefont{Xu}},
  \bibinfo{author}{\bibfnamefont{H.}~\bibnamefont{Jin}},
  \bibinfo{author}{\bibfnamefont{L.}~\bibnamefont{Zhang}},
  \bibinfo{author}{\bibfnamefont{W.}~\bibnamefont{Peng}},
  \bibinfo{author}{\bibfnamefont{Z.}~\bibnamefont{Wang}}, \bibnamefont{et~al.},
  \bibinfo{journal}{unpublished}  (\bibinfo{year}{2022}{\natexlab{b}}).

\bibitem[{\citenamefont{Kramer and MacKinnon}(1993)}]{Kramer_1993}
\bibinfo{author}{\bibfnamefont{B.}~\bibnamefont{Kramer}} \bibnamefont{and}
  \bibinfo{author}{\bibfnamefont{A.}~\bibnamefont{MacKinnon}},
  \bibinfo{journal}{Rep. Prog. Phys.} \textbf{\bibinfo{volume}{56}},
  \bibinfo{pages}{1469} (\bibinfo{year}{1993}).

\bibitem[{\citenamefont{Oya and Onodera}(1974)}]{Oya1974}
\bibinfo{author}{\bibfnamefont{G.}~\bibnamefont{Oya}} \bibnamefont{and}
  \bibinfo{author}{\bibfnamefont{Y.}~\bibnamefont{Onodera}},
  \bibinfo{journal}{J. Appl. Phys.} \textbf{\bibinfo{volume}{45}},
  \bibinfo{pages}{1389} (\bibinfo{year}{1974}).

\bibitem[{\citenamefont{Wang et~al.}(1996)\citenamefont{Wang, Kawakami, Uzawa,
  and Komiyama}}]{Wang1996}
\bibinfo{author}{\bibfnamefont{Z.}~\bibnamefont{Wang}},
  \bibinfo{author}{\bibfnamefont{A.}~\bibnamefont{Kawakami}},
  \bibinfo{author}{\bibfnamefont{Y.}~\bibnamefont{Uzawa}}, \bibnamefont{and}
  \bibinfo{author}{\bibfnamefont{B.}~\bibnamefont{Komiyama}},
  \bibinfo{journal}{J. Appl. Phys.} \textbf{\bibinfo{volume}{79}},
  \bibinfo{pages}{7837} (\bibinfo{year}{1996}).

\end{thebibliography}


\end{document}